\documentclass[12pt]{iopart}
\usepackage{psfig}
\begin{document}

\title{Atomic entanglement near a realistic microsphere}

\author{Ho Trung Dung, S Scheel, D-G Welsch and L Kn\"oll}
\address{
Theoretisch-Physikalisches Institut, 
Friedrich-Schiller-Universit\"{a}t Jena, 
Max-Wien-Platz 1, 07743 Jena, Germany}

\begin{abstract}
We study a scheme for entangling two-level atoms located close to
the surface of a dielectric microsphere. The effect is
based on medium-assisted spontaneous decay, rigorously taking
into account dispersive and absorptive properties of the microsphere.
We show that even in the weak-coupling regime, where the Markov
approximation applies, entanglement up to $0.35$ ebits between two atoms
can be created. However, larger entanglement and violation of
Bell's inequality can only be achieved in the strong-coupling regime.
\end{abstract}

\submitto{\JOB}
\maketitle


\section{Introduction}
\label{Intro}

A bipartite quantum system is said to be entangled if its state cannot be 
represented as a convex sum of direct products of its subsystem states.
Quantum entanglement entails correlations between outcomes
of particular experiments at (spatially) separated objects which 
break certain Bell's inequalities, predicted by local realistic
theories \cite{Bell65}. Many experiments have been performed 
to test Bell's inequalities \cite{Fry00}, with the locality \cite{Weihs98}
and detection efficiency \cite{Rowe01} loopholes recently reported to
be closed. Despite this breakthrough, a decisive experiment to rule
out any local realistic theory is yet to be performed \cite{Vaidman01}. 
Beyond the fundamental aspects, entanglement is a key resource
for many applications in quantum information processing,
including secret key distribution \cite{Ekert91}, 
dense coding \cite{Bennett92}, and teleportation \cite{Bennett93}.

Atoms can be entangled through interaction with a (common) electromagnetic
field. The effect, which is very weak in free space, can be enhanced 
significantly in resonator-like equipments. Proposals have been 
made for entangling spatially separated atoms in Jaynes-Cummings 
systems through strong resonant atom-field coupling \cite{Kudryavtsev93}. 
The coupling can be sequential or simultaneous.
One of these schemes has been realized 
using Rydberg atoms coupled one by one to a high $Q$ 
microwave superconducting microcavity \cite{Hagley97}, with
achieved probability of preparing a maximally entangled state in the 
range of $0.63$ and two atoms separated by centimetric distances. 
To limit photon losses, off-resonant coupling, where the cavity mode is 
only virtually excited, can also be employed as recently 
proposed \cite{Zheng00} and implemented \cite{Osnaghi01}.
Another proposal involves continuous monitoring of photons 
leaking out of the cavity. Provided no photon is detected outside the 
cavity, a pure entangled state between two atoms results 
\cite{Plenio99,Beige00}.

In this paper we first consider the more general problem of generation
of two-atom entangled states in the presence of dispersing and
absorbing dielectric bodies owing to the medium-assisted change of
the spontaneous emission and the mutual dipole-dipole coupling of
the atoms. Our investigation is based on a macroscopic approach to
the electromagnetic field quantization which represents the potential
and field operators in terms of a Green tensor expansion over 
appropriately chosen fundamental bosonic field variables  
(for a review, see \cite{Knoll01}).

We apply the theory to the case of the two atoms being near a
microsphere and show that the scheme, in contrast to much of
previous work, does not require a strong atom-field coupling
regime to realize entanglement, but the resulting state is not
maximally entangled. Further, we study the strong-coupling
regime, taking rigorously into account atomic spontaneous decay, 
photon leakage from the microsphere, and material absorption and
dispersion. As an example, numerical calculations are performed
for a microsphere whose permittivity is of Drude--Lorentz type.


\section{Basic equations}
\label{sec:basic}

Let us consider $N$ two-level atoms [positions ${\bi r}_A$, transition
frequencies $\omega_A$ ($A$ $\!=$ $\!1,2,...,N$)] 
that resonantly interact with radiation
via electric-dipole transitions (dipole moments ${\bi d}_A$).
Let us further assume that the atoms are sufficiently far from each
other, so that the interatomic Coulomb interaction can be ignored.
In this case, the electric-dipole and rotating
wave approximations apply, and the minimal-coupling 
Hamiltonian takes the form of \cite{Knoll01}
\begin{eqnarray}
\label{e3.1}
\fl
   \hat{H} = \int \!{\rm d}^3{\bi r}
   \! \int_0^\infty \!\!{\rm d}\omega \,\hbar\omega
   \,\hat{\bi f}^\dagger({\bi r},\omega){}\hat{\bi f}({\bi r},\omega)
    + \sum_A {\textstyle{1\over 2}}\hbar\omega_A \hat{\sigma}_{Az}
    - \sum_A \big[
    \hat{\sigma}_A^\dagger
    \hat{\bi E}^{(+)}({\bi r}_A){}{\bi d}_A  
    + {\rm H.c.}\big],
\end{eqnarray}
where the two-level atoms are described
by the Pauli operators $\hat{\sigma}_A$,
$\hat{\sigma}_A^\dagger$, and $\hat{\sigma}_{Az}$,
and $\hat{\bi f}({\bi r},\omega)$ and
$\hat{\bi f}^\dagger({\bi r},\omega)$ are bosonic field operators
which play the role of the fundamental variables of the
electromagnetic field and the medium, including a reservoir 
necessarily associated with the losses in the medium.
The electric-field operator is expressed in terms of 
$\hat{\bi f}({\bi r},\omega)$ as
\begin{eqnarray}
\label{e2.2}
     \hat{\bi E}^{(+)}({\bi r}) 
      = i \sqrt{\frac{\hbar}{\pi\varepsilon_0}}
     \int_{0}^{\infty} \!{\rm d}\omega \,\frac{\omega^2}{c^2}
   \!\!\int \!\!{\rm d}^3{\bi r}'
   \sqrt{\varepsilon_{\rm I}({\bi r}',\omega)}
   \,{\bf G}({\bi r},{\bi r}',\omega)   
   {}\hat{\bi f}({\bi r}',\omega),
\end{eqnarray}
where ${\bf G}({\bi r},{\bi r}',\omega)$ is the classical
Green tensor and \mbox{$\varepsilon({\bi r},\omega)$
$\!=$ $\!\varepsilon_{\rm R}({\bi r},\omega)$
$\!+$ $\!i\varepsilon_{\rm I}({\bi r},\omega)$} the
complex permittivity.

For a single-quantum excitation, the system wave 
function at time $t$ can be written as
\begin{eqnarray}
\label{e3.2}
\lefteqn{
    |\psi(t)\rangle = \sum_A C_{U_A}(t)
    \rme^{-\rmi(\omega_A-\bar{\omega})t}
    |U_A\rangle |\{0\}\rangle
}
\nonumber\\[.5ex]&&\hspace{8ex}
     + \int {\rm d}^3{\bi r} \int_0^\infty {\rm d}\omega\, 
     \Bigl[ C_{Li}({\bi r},\omega,t)
     \rme^{-\rmi (\omega-\bar{\omega})t}
     |L\rangle |\{1_i({\bi r},\omega)\} \rangle \Bigr]
\end{eqnarray}
($\bar{\omega}$ $\!=$ $\!$ $\!\frac{1}{2}\sum_A\omega_A$).
Here, $|U_A\rangle$ is the excited atomic state, where
the $A$th
atom is in the upper state and all the other
atoms are in the lower state, and $|L\rangle$ is the
atomic state, where all atoms are in the lower state.
Accordingly, $|\{0\}\rangle$ is the vacuum state of the 
rest of the system, and $|\{1_i({\bi r},\omega)\}\rangle$ 
$\!=$ $\!\hat{f}^\dagger_i({\bi r},\omega) |\{0\}\rangle$
is the state, where it is excited in a single-quantum Fock state.
Note that this state is not a photonic state in general, but a state
of the macroscopic medium dressed by the electromagnetic field.
From the Schr\"odinger equation, we obtain the 
following (closed) system of integro-differential equations:
\begin{eqnarray}
\label{e8.5}
\fl
        \dot{C}_{U_A}(t) = \sum_{A'}
        \int_0^t {\rm d}t'\, K_{AA'}(t,t')\, C_{U_{A'}}(t')
        + f_A(t),
\\[.5ex]
\label{e8.5a}
\fl
   f_A(t) =
   -\frac{1}{\sqrt{\pi\varepsilon_0\hbar}}           
   \int_0^\infty \!{\rm d}\omega
   \int{\rm d}^3{\bi r}\,
   \bigg\{
   \frac{\omega^2}{c^2}\,\rme^{-\rmi(\omega-\omega_A)t}  
   \Big[ \sqrt{\varepsilon_{\rm I}({\bi r},\omega)}
   \,{\bi d}_A{\bf G}({\bi r}_A,{\bi r},\omega)
   {\bi C}_{L}({\bi r},\omega,0) \Big]
   \bigg\},
\\[.5ex]
\label{e8.6} 
\fl
        K_{AA'}(t,t') =
        -\frac{1}
          {\hbar\pi\varepsilon_0}
        \int_0^\infty {\rm d}\omega \, \biggl[
        {\omega^2\over c^2}
        \rme^{-\rmi(\omega-\omega_A)t}
        \rme^{\rmi(\omega-\omega_{A'})t'}         
       {\bi d}_A {\rm Im}\,{\bf G}({\bi r}_A,{\bi r}_{A'},\omega)
       {\bi d}_{A'} \biggr].
\qquad
\end{eqnarray}
Note that $K_{AA'}(t,t')=K_{A'A}^\ast(t',t)$
because of the reciprocity theorem.

The excitation can initially reside either in
one of the atoms or the medium-assisted electromagnetic field.
The latter case, i.e., 
\mbox{${\bi C}_L({\bi r},\omega,0)$ $\!\neq$
$\!0$} in equation~(\ref{e8.5a}), could be realized, for example, 
by coupling the field first to an excited atom {$D$}
during a time interval $\Delta t$ such that
\begin{eqnarray}
\label{e3.35}
\fl
   {\bi C}_{L}({\bi r},\omega,0) =  \int_{-\Delta t}^0 {\rm d}t'\,
   \frac{1}{\sqrt{\pi\varepsilon_0\hbar}}
   \,\frac{\omega^2}{c^2}
   \,\rme^{\rmi(\omega-\omega_D)t'}
   \sqrt{\varepsilon_{\rm I}({\bi r},\omega)}          
   \,{\bi d}_D{\bf G}^\ast({\bi r}_D,{\bi r},\omega)\, C_{U_D}(t'),
\end{eqnarray}
where $C_{U_D}(t)$
describes the single-atom decay \cite{Ho00}. Substitution of the
expression (\ref{e3.35}) into equation~(\ref{e8.5a}) then yields
\begin{eqnarray}
\label{e8.6b}
        f_A(t)= \int_{-\Delta t}^0 {\rm d}t'\, K_{AD}(t,t')\,C_{U_D}(t').
\end{eqnarray}

We now turn to the problem of two atoms, denoted by $A$ and $B$.
For simplicity, let us consider atoms with equal transition frequencies
$\omega_A$ $\!=$ $\!\omega_B$ $\!=$ $\!\omega_D$,
so that
\begin{equation}
\label{e8.20b}
       K_{AA'}(t,t') \equiv K_{AA'}(t-t')
\end{equation}
($A'$ $\!=$ $\!B,D$) and
\begin{equation}
\label{e8.20c}
       K_{AB}(t-t') = K_{BA}(t-t'),
\end{equation}
and assume that the isolated atoms 
follow the same decay law,
\begin{equation}
\label{e8.20a}
   K_{AA}(t,t') =  K_{BB}(t,t') \equiv K(t-t'). 
\end{equation}
Introducing the new variables
\begin{equation}
\label{e8.12}
      C_\pm (t) = 2^{-1/2}
      [ C_{U_A}(t) \pm C_{U_B}(t) ],
\end{equation}
it is not difficult to prove that the system of equations (\ref{e8.5})
[together with equation (\ref{e8.6b})] decouples as follows:
\begin{eqnarray}
\label{e8.17}
\fl
       \dot{C}_\pm(t) = \int_{0}^t {\rm d}t'\, K_\pm(t-t')\,C_\pm(t')
       + {1\over\sqrt{2}}
       \int_{-\Delta t}^0 \!\!{\rm d}t'
       \left[K_{AD}(t-t')
       \pm K_{BD}(t-t')\right]C_{U_D}(t') ,
\end{eqnarray}
where
\begin{equation}
\label{e8.18}
       K_\pm(t-t') = K(t-t') \pm  K_{AB}(t-t').
\end{equation}
Obviously, the $C_{\pm}(t)$ are the expansion coefficients
of the wave function with respect to the (atomic) basis 
\begin{equation}
\label{e8.14}
      |\pm \rangle = 2^{-1/2} 
      \left( |U_A\rangle \pm |U_B\rangle \right)
\end{equation}
and $|L\rangle$ (instead of the basis $|U_A\rangle$,
$|U_B\rangle$, and $|L\rangle$). Thus, $C_+(t)$ and $C_-(t)$
are the probability amplitudes of finding the total system in the 
states  $|+\rangle |\{0\}\rangle$ and $|-\rangle |\{0\}\rangle$, 
respectively. In the further treatment of equation~(\ref{e8.17})
one can distinguish between the weak- and the strong-coupling
regime.


\subsection{Weak Coupling}

In the weak-coupling regime, the Markov approximation applies,
and in equation~(\ref{e8.17}) $C_\pm(t')$ 
can be replaced with $C_\pm(t)$,
with the time integrals being $\zeta$-functions.
In particular, when the medium-assisted field is initially not excited,
then the second term on the right-hand side of equation~(\ref{e8.17})
vanishes and we are left with a homogeneous first-order differential 
equation, whose solution is
\begin{equation}
\label{e8.13}
      C_\pm (t)
       = \rme^{(-\Gamma_\pm/2 +\rmi \delta_\pm) t} C_\pm(0),
\end{equation}
where ($\Gamma$ $\!\equiv$ $\Gamma_{AA}$,
$\delta$ $\!\equiv$ $\delta_{AA}$)
\begin{equation}
\label{e8.9}
      \Gamma_\pm = \Gamma \pm \Gamma_{AB}\,,
      \qquad
      \delta_\pm = \delta \pm \delta_{AB}\,, 
\end{equation}
\begin{equation}
\label{e8.10} 
     \Gamma_{AB} = {2\omega_A^2\over \hbar\varepsilon_0c^2}\, 
     {\bi d}_A {\rm Im}\,{\bf G}({\bi r}_A,{\bi r}_B,\omega_A) {\bi d}_B \,,
\end{equation}
\begin{equation}
\label{e8.11} 
   \delta_{AB} \!=\! {{\cal P}\over \pi\hbar\varepsilon_0}\,
   \!\int_0^\infty \!\!{\rm d}\omega\,
   {\omega^2\over c^2} 
   \frac{{\bi d}_A {\rm Im}\,{\bf G}({\bi r}_A,{\bi r}_B,\omega) {\bi d}_B}
   {\omega-\omega_A}\,. 
\end{equation}
Obviously, $\Gamma_\pm$
are the decay rates of the states $|\pm\rangle$, and  
the assumption (\ref{e8.20a}) means
that the two atoms are positioned in such a way that they have 
equal single-atom decay rates and line shifts.
Note that the values of $\Gamma_+$ and $\Gamma_-$ can
substantially differ from each other, because of the
interference term $\Gamma_{AB}$ (of positive or negative sign).   

At this point it should be mentioned that, on starting from the
Hamiltonian (\ref{e3.1}), the reduced density operator for the
atomic subsystem in the weak-coupling regime can be shown to
obey the equation
\begin{eqnarray}
\label{e8.11a}
\fl
      \dot{\hat{\varrho}} =
      &&-{\textstyle{1\over 2}}\sum_{A,A'} \Gamma_{AA'} \left(
      \hat{\sigma}^\dagger_A \hat{\sigma}_{A'} \hat{\varrho}
      -2\hat{\sigma}_{A'} \hat{\varrho} \hat{\sigma}^\dagger_A
      +\hat{\varrho}\hat{\sigma}^\dagger_A \hat{\sigma}_{A'} 
      \right)
      +\rmi\sum_{A,A'} \delta_{AA'}
      [\hat{\sigma}^\dagger_A \hat{\sigma}_{A'}, \hat{\varrho}]
\end{eqnarray}
($A,A'$ $\!=$ $\!1,2,...,N$).
It is worth noting that this result is in agreement
with the result given in \cite{Agarwal75}, where Kubo's
formula is applied to the field correlation functions.


\subsection{Strong Coupling}

In the strong-coupling regime (i.e., when the atoms are in
a resonator-like equipment of sufficiently high quality),
the atoms are predominantly coupled  
to a sharp field resonance, whose mid-frequency approximately
equals the atomic transition frequency. As a result, the probability
amplitudes in equation~(\ref{e8.17}) must not necessarily
be slowly varying compared with the kernel functions and
the Markov approximation thus fails in general.  
Regarding the line shape of the field resonance as being
a Lorentzian, one can of course approximate the kernels
$K(t-t')$, $K_{AB}(t-t')$ [and $K_{AD}(t-t')$ and $K_{BD}(t-t')$]
in a similar way as done in \cite{Ho00} for a single atom.

Equation (\ref{e8.17}) reveals that the motion of the
states $|\pm\rangle$ defined by equation~(\ref{e8.14})
is governed by the kernel functions
$K_\pm(t-t')$, and it may happen that one of them becomes
very small, because of destructive interference 
[cf. equation~(\ref{e8.18})]. In that case, either
$|+\rangle$ or $|-\rangle$ is weakly coupled to the
field, and thus  
the strong-coupling regime cannot be realized for
both of these states simultaneously.

\section{Entangled-state preparation}
\label{subsec:ent_gen}

Let us consider a particular configuration of  
material bodies, namely, a microsphere (radius $R$), which can act 
as a microcavity \cite{Chang96}. It is well known that rays
of suitable wavelengths may bounce around the rim by total internal 
reflection, forming the so-called whispering gallery (WG) 
waves. These field resonances can combine extreme 
photonic confinement with very high quality factors 
-- properties that are crucial for cavity QED experiments \cite{Lin92}
and many optoelectronical applications \cite{Chang96}.
For a band-gap material and frequencies inside the band-gap,
a microsphere can also give rise to high quality surface-guided
(SG) waves \cite{Ho01b}.

Here we assume that the microsphere material can be
characterized by a (single-resonance) Drude--Lorentz-type
permittivity
\begin{equation}
\label{e4.5}
        \varepsilon(\omega) = 1 + 
        {\omega_{\rm P}^2 \over 
        \omega_{\rm T}^2 - \omega^2 - i\omega \gamma}\,,
\end{equation}
where $\omega_{\rm P}$ corresponds to the coupling constant, 
and $\omega_{\rm T}$ and $\gamma$ are the medium
oscillation frequency and the linewidth, respectively. Recall that the
Drude--Lorentz model covers both metallic
(\mbox{$\omega_{\rm T}=0$}) and 
dielectric (\mbox{$\omega_{\rm T}\neq0$}) matter 
and features a band gap between $\omega_{\rm T}$ and 
\mbox{$\omega_{\rm L}$ $\!=$ 
$\!\sqrt{\omega_{\rm T}^2+\omega_{\rm P}^2}$}.

\subsection{Weak Coupling}

Let us restrict our attention to two identical (two-level) atoms
located at diametrically opposite positions 
(\mbox{${\bi r}_A$ $\!=$ $\!-{\bi r}_B$}) outside the microsphere 
and having radially oriented transition dipole moments.
Obviously, the conditions (\ref{e8.20c}) and (\ref{e8.20a})
are fulfilled for such a system, so that from equations~(\ref{e8.9})
and (\ref{e8.10}) together with the Green tensor for a
microsphere \cite{Li94} one can find that
\begin{eqnarray}
\label{e8.16}
\lefteqn{
      \Gamma_\pm = {\textstyle{3\over 2}} \Gamma_0
      \sum_{l=1}^\infty
      {\rm Re}\biggl\{
      {l(l+1)(2l+1) \over (k_Ar_A)^2}      
      h^{(1)}_l(k_Ar_A) 
}
\nonumber\\[.5ex]&&\hspace{5ex} \times
      \left[ j_l(k_Ar_A)+ {\cal B}^N_l\!(\omega_A)
      h^{(1)}_l(k_Ar_A) \right] 
      \left[ 1\mp(-1)^l\right]
      \biggr\}
\end{eqnarray}
[$k_A$ $\!=$ $\omega_A/c$;  
$j_l(z)$ and $h^{(1)}_l(z)$, spherical Bessel and 
Hankel functions; ${\cal B}^N_l\!(\omega_A)$, generalized 
reflection coefficient \cite{Ho01b}; 
$\Gamma_0$, decay rate of a single atom in free space].
When atom $A$ is initially in the
upper state and atom $B$ is accordingly in the lower state, then 
the two superposition states
$|+\rangle$ and $|-\rangle$ [equation~(\ref{e8.14})] are 
initially equally excited [\mbox{$C_+(0)$ $\!=$ $\!C_-(0)$ $\!=$
$\!2^{-1/2}$}].
If the atomic transition frequency coincides with a microsphere
resonance, the single-atom decay rate $\Gamma$ may be approximated
(for sufficiently small atom-surface distance) by \cite{Ho01b}
\begin{equation}
\label{e8.16b}
\Gamma
   \simeq {\textstyle{3\over 2}} \Gamma_0\,
      l(l+1)(2l+1)
      {\rm Re} \biggl\{
      \biggl[{h^{(1)}_l(k_Ar_A) \over k_Ar_A}\biggr]^2
      {\cal B}^N_l\!(\omega_A) 
      \biggr\},
\end{equation}
and equation~(\ref{e8.16}) can be approximated as follows:
\begin{equation}
\label{e8.16c}
\Gamma_\pm \simeq \Gamma \left[1 \mp (-1)^l\right]. 
\end{equation}
Hence \mbox{$\Gamma_-$ $\!\gg$ $\!\Gamma_+$}
(\mbox{$\Gamma_+$ $\!\gg$ $\!\Gamma_-$})
if $l$ is even (odd), i.e., the state $|-\rangle$ ($|+\rangle$)
decays much faster than the state $|+\rangle$ ($|-\rangle$).

\begin{figure}[ht]
\hspace*{2cm}
\psfig{figure=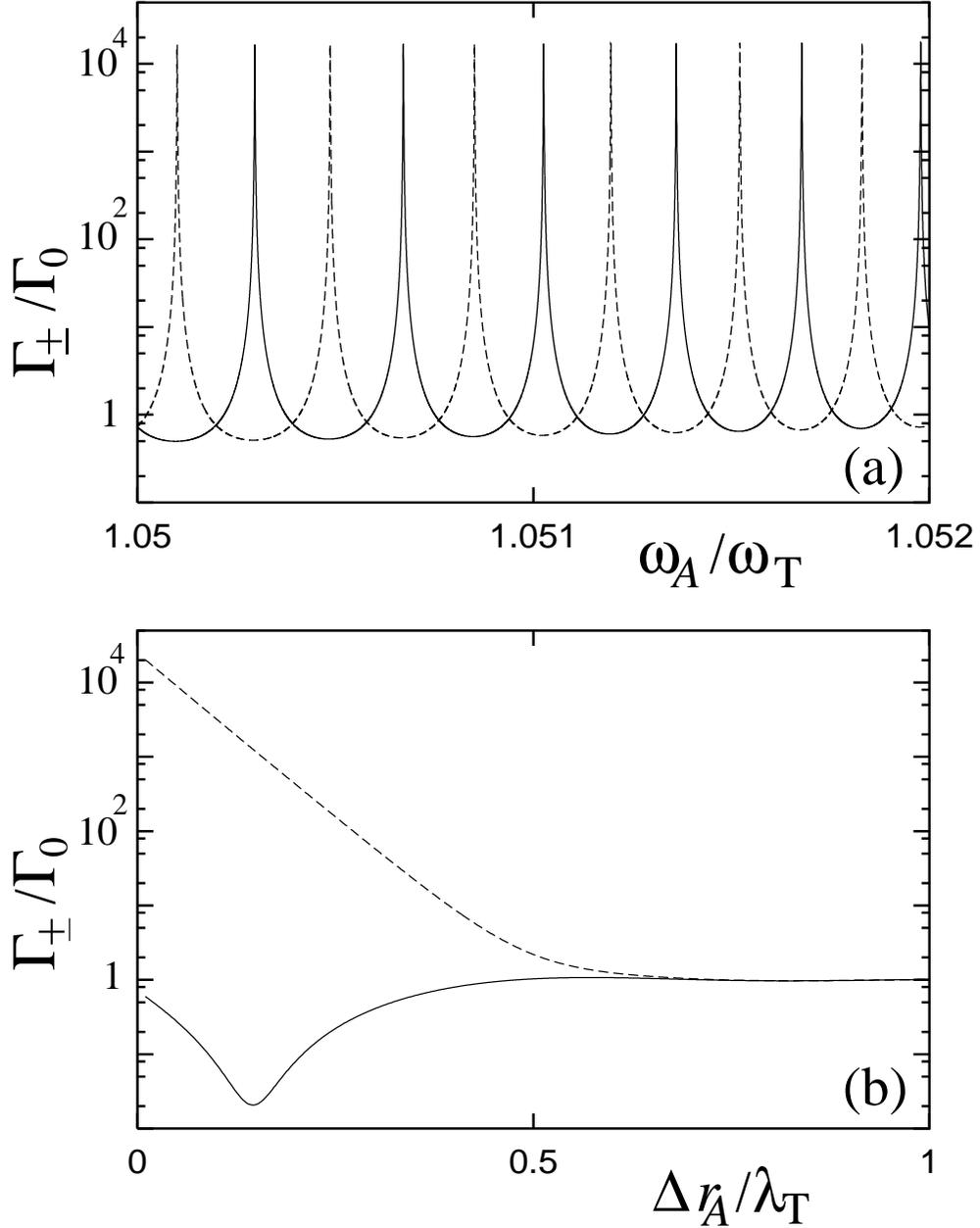,width=.85\textwidth}
\caption{
The dependence of the decay rates $\Gamma_+$ (solid line)
and $\Gamma_-$ (dashed line) on (a) the transition
frequency and (b) the distance of the atoms from
a microsphere is shown for two atoms at (with respect
to a sphere) diametrically
opposite positions, radially oriented
transition dipole moments, and a single-resonance
Drude--Lorentz-type dielectric
[\mbox{$R$ $\!=$ $\!10\,\lambda_{\rm T}$
\mbox{($\lambda_{\rm T}$ $\!=$ $\!2\pi c/\omega_{\rm T}$)}};
\mbox{$\omega_{\rm P}$ $\!=$ $\!0.5\,\omega_{\rm T}$};
\mbox{$\gamma$ $\!=$ $\!10^{-6}\omega_{\rm T}$};
\mbox{$\Delta r_B$ $\!\equiv$ $\!r_B$ $\!-$ $\!R$ 
$\!=$ $\!\Delta r_A$
$\!\ge$ $\!10^{-2}\,\lambda_{\rm T}$};
(a) \mbox{$\Delta r_A$ $\!=$ $\!0.02\,\lambda_{\rm T}$};
(b) \mbox{$\omega_A$ $\!\simeq$ $\!1.0501\,\omega_{\rm T}$}].
}
\label{fig:ent}
\end{figure}
The (exact) frequency dependence of $\Gamma_\pm$ as given by
equation~(\ref{e8.16}) is illustrated in figure~\ref{fig:ent}(a)
for a frequency interval inside a band gap, and the
dependence on the atom-surface distance is illustrated
in figure~\ref{fig:ent}(b). 
We see that the values of $\Gamma_+$ and $\Gamma_-$ can
be substantially different from each other before they
tend to the free-space rate $\Gamma_0$ as the distance
from the sphere becomes sufficiently large. In particular, the
decay of one of the states $|+\rangle$ or $|-\rangle$ can
strongly be suppressed at the expense
of the other one, which decays rapidly.  
Note that $\Gamma_+$ also differs from $\Gamma_-$ for
two atoms in free space \cite{DeVoe96}. However,
the difference that occurs by mediation of the microsphere
is much larger. For example, at the distance for
which in figure~\ref{fig:ent}(b) $\Gamma_+$ attains the
minimum, the ratio 
\mbox{$\Gamma_-/\Gamma_+$ $\!\simeq$ $\!6.7\!\times\!10^4$} is
observed, which is to be compared with the free-space
ratio \mbox{$\Gamma_-/\Gamma_+$ $\!\simeq$ $\!1.0005$}.
The effect may become even more pronounced for larger microsphere
sizes and lower material absorption, i.e., sharper microsphere
resonances. Needless to say that it is not only
observed for SG waves considered in
figure~\ref{fig:ent}, but also for WG waves.  

{F}rom the above, there exists a time window,
during which the overall system is prepared in an entangled state
that is a superposition of the state with the atoms being
in the state $|+\rangle$ ($|-\rangle$)
and the medium-assisted field
being in the ground state, and all the states
with the atoms being in the lower state $|L\rangle$
and the medium-assisted field being in a single-quantum
Fock state. The window is opened
when the state $|-\rangle$ ($|+\rangle$) has already decayed
while the state $|L\rangle$ emerges, and
it is closed roughly after the lifetime of the state $|+\rangle$
($|-\rangle$). 
As a result, the two atoms are also entangled to each other.
The state is a statistical mixture,
the density operator of which is
obtained from the density operator of 
the overall system by taking the trace with respect
to the medium-assisted field. Within the approximations 
(\ref{e8.13}) and (\ref{e8.16c}) it takes the form of
\begin{eqnarray}
\label{e8.16a}
      \hat{\varrho} \simeq 
      |C_\pm(t)|^2 |\pm\rangle\langle\pm|
      + \left[1-|C_\pm(t)|^2 \right] 
        |L\rangle\langle L|,
\end{eqnarray}
where
\begin{equation}
\label{e8.16d}
|C_\pm(t)|^2 \simeq 2^{-1} \rme^{-\Gamma_\pm t} .
\end{equation}

Applying the separability criterion \cite{Peres96},
it is not difficult to prove that the
state (\ref{e8.16a}) is indeed inseparable, in fact, for all times $t$.
It is worth noting that the atoms become entangled 
within the weak-coupling regime, starting from
the state $|U_A\rangle$ (or $|U_B\rangle$) and the
vacuum field. In the language of (Markovian)
damping theory one would probably say that the two atoms are
coupled to the same dissipative system, which gives rise to
the quantum coherence. 

\begin{figure}[ht]
\hspace*{2.2cm}
\psfig{figure=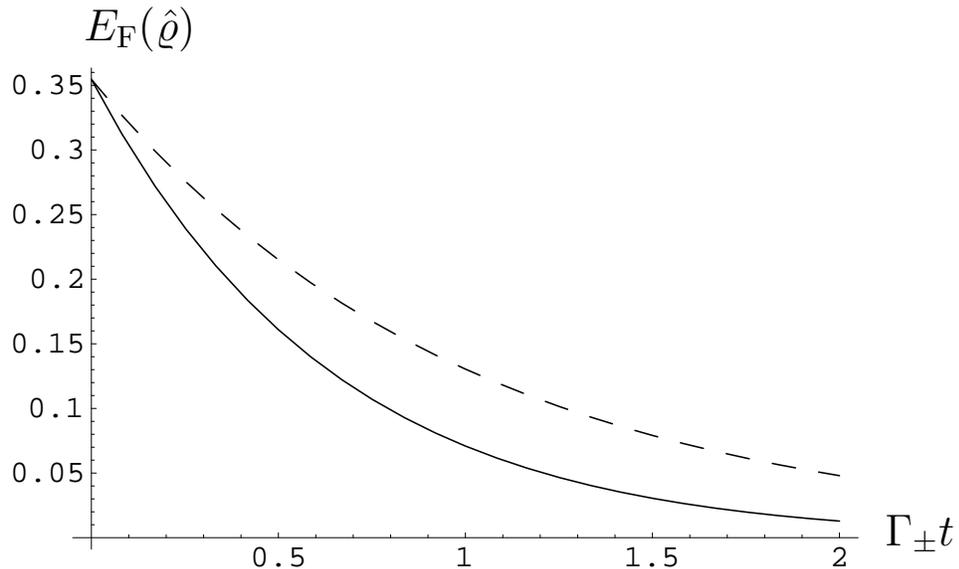,width=.85\textwidth}
\caption{\label{fig:schwach}
Entanglement of formation in the weak-coupling regime
(\protect$\Gamma_\pm$ $\!\ll$ $\!\Gamma_\mp$). For comparison,
$2E_{\rm F}(\hat{\varrho})|_{t\to0}|C_\pm(t)|^2$ is shown (dashed
line).}
\end{figure}

The time evolution of the entanglement of formation
$E_{\rm F}(\hat{\varrho})$ (for the concept of entanglement
of formation, see \cite{Hill97}) is shown in
figure~\ref{fig:schwach}, where
the entanglement is measured in ebits.
%
It is clear from the structure of the coefficient $C_\pm(t)$
[equation~(\ref{e8.16d})] that one can never achieve a maximally entangled
state in the weak-coupling regime, since the state (\ref{e8.16a}) 
is never pure. Applying the
convexity property of entanglement measures, one realizes that the
amount of entanglement contained in the state (\ref{e8.16a}) is
bounded according to \cite{Scheel00}
\begin{equation}
E_{\rm F}(\hat{\varrho}) \le |C_\pm(t)|^2.
\end{equation}
Figure \ref{fig:schwach} reveals that at most $0.35$~ebits
can be achieved for \mbox{$t$ $\!\to$ $\!0$} (the limit
\mbox{$t\to 0$} is to
be understood as the smallest time that is compatible with the
requirement for the fast decay channel to be already closed).
It is further seen that the entanglement falls
off faster than exponentially with time -- a result that is
already expected from the convexity property and
equation~(\ref{e8.16d}).

Entangled-state preparation in the weak-coupling re\-gime
has the advantage that it could routinely be achieved experimentally. 
However, the value of $|C_\pm(t)|^2$ in equation~(\ref{e8.16a})
is always less than $1/2$. In order to achieve a higher
degree of ent\-angle\-ment, a strong-coupling regime is required.


\subsection{Strong Coupling}

Let us assume that the two atoms are initially in the
ground state and the medium-assisted field is excited.
The field excitation can be achieved, for example,
by coupling an excited atom $D$ to the microsphere and
then making sure that the atomic excitation is transferred to 
the medium-assisted field. If the atom $D$ strongly interacts
with the field, the excitation transfer can be controlled  
by adjusting the interaction time.
Another possibility would be measuring the state populations
and discarding the events where the atom
is found in the upper state. Here we restrict our attention
to the first method and assume that all three
atoms $D$, $A$, and $B$ strongly interact with the same
microsphere resonance (of mid-frequency $\omega_{\rm C}$ and line
width $\Delta\omega_{\rm C}$). The upper-state probability
amplitude $C_{U_D}(t)$ of atom $D$ can then given by \cite{Ho00}
\begin{equation}
\label{e8.19} 
       C_{U_D}(t) = \rme^{-\Delta\omega_{\rm C} (t+\Delta t)/2}
       \cos [\Omega_D (t+\Delta t)/2], 
\end{equation}
with $\Omega_D$ being the corresponding Rabi frequency
\mbox{$\Omega$ $\!=$ $\!\sqrt{2\Gamma_{\rm C}\Delta\omega_{\rm C}}$},
where $\Gamma_{\rm C}$ is the value of $\Gamma$ at the frequency
$\omega_{\rm C}$; for the calculation of $\Delta\omega_{\rm C}$,
see \cite{Ho01b}. For
\begin{equation}
\label{e8.22}
       \Delta t =  \pi/\Omega_D ,
\end{equation}
the initially (i.e., at time \mbox{$t$ $\!=$ $\!-\Delta t$}) excited
atom $D$ is at time $t$ $\!=$ $\!0$ in the lower state
[\mbox{$C_{U_D}(0)$ $\!=$ $\!0$}].

{F}rom the preceding subsection we know that
when the resonance angular-momentum number $l$ is odd (even),
then the state $|+\rangle$ ($|-\rangle$) ``feels'' a sharply
peaked high density of medium-assisted field states, so that a 
strong-coupling approximation applies.
The state $|-\rangle$ ($|+\rangle$), in contrast, ``feels'' a flat
one and the (weak-coupling) Markov approximation applies.
Assuming atom $A$ is at the same position
as was atom $D$, from equations~(\ref{e8.17}), (\ref{e8.19}),
and (\ref{e8.22}) we then find that
\begin{equation}
\label{e8.21}
       C_\pm(t) \simeq - \rme^{-\Delta\omega_{\rm C}( t +\pi/\Omega_D)/2}
          \sin(\Omega_\pm t/2)
\end{equation}
($\Omega_\pm$ $\!=$ $\!\sqrt{2}\Omega$, 
with $\Omega$ being the Rabi frequency
of atom $A$ or $B$), and
\begin{equation}          
\label{e8.21a} 
       C_\mp(t) \simeq 0
\end{equation}
(the sign of $C_-(t)$ in equation~(\ref{e8.21}) is reversed
if atom $B$ is at the same position as was atom $D$).
The two-atom entangled state is again of the form~(\ref{e8.16a}),
but now the weight of the
state $|+\rangle$ ($|-\rangle$) can reach values larger $1/2$,
provided that the resonance linewidth $\Delta\omega_{\rm C}$ is small enough.
%
\begin{figure}[ht]
\hspace*{-2.5cm}
\psfig{figure=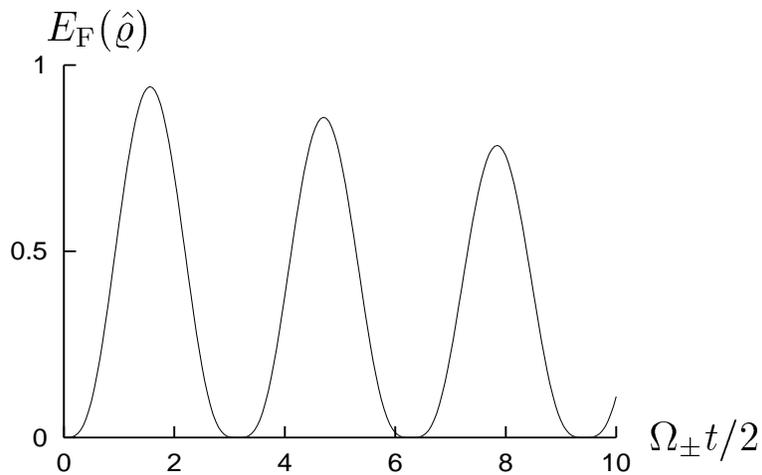,width=1.\textwidth}
\caption{\label{fig:stark} Entanglement of formation in the
strong-coupling regime (\protect$\Delta\omega_{\rm C}/\Omega_\pm$
$\!=$ $\!0.01$; $\pi\Delta\omega_{\rm C}/\Omega_D$ $\!=$ $\!0.01$).}
\end{figure}
An example of the time evolution of the entanglement of formation
is shown in figure~\ref{fig:stark}.
The shape of the curve strongly depends on the
ratios $\Delta\omega_{\rm C}/\Omega_\pm$ and $\Delta\omega_{\rm C}/\Omega_D$.
Small values of $\Delta\omega_{\rm C}/\Omega_\pm$ yield many oscillations with,
on assuming that $\pi\Delta\omega_{\rm C}/\Omega_D$ is also small, large
achievable entanglement. This is the case in figure~\ref{fig:stark}.
Roughly speaking, 
$\Omega_D$ controls the maximal obtainable entanglement,
$\Delta\omega_{\rm C}$ the decrease of the envelope, and $\Omega_\pm$ the
oscillation frequency. Highest entanglement is achieved if
\begin{equation}
\label{cond}
\Delta\omega_{\rm C} \ll \Omega_\pm,\,\Omega_D.
\end{equation}
For example, choose a microsphere with a $Q$ factor of
$2.4\!\times\!10^8$. At
optical frequencies of \mbox{$\omega_{\rm C}\sim 3\!\times\!10^{15}\,$}Hz
(i.e., \mbox{$\lambda\approx 600$\,}nm) this amounts to 
\mbox{$\Delta\omega_{\rm C}\sim 10^7\,$}Hz.
For \mbox{$\lambda\approx 573$\,}nm, the spontaneous decay rate of a
quantum dot, regarded as an artificial atom, has been measured to be
\mbox{$\Gamma\approx 5\!\times\!10^9$\,}Hz
\cite{Fan00}, so that
it follows that \mbox{$\Omega\sim 10^1\ldots 10^2\Delta\omega_{\rm C}$}.
This justifies the parameters chosen in figure~\ref{fig:stark}.
The results also show that maximal entanglement of $1\,$ebit cannot
be achieved in practice even in the strong-coupling regime.


\subsection{Multiparticle entanglement}

Besides entangling two atoms, the scheme can also be used to create
multiparticle entanglement.
Let us briefly discuss the problem of creating 
tripartite entanglement by a single excitation.
Then, instead of the states $|U_A\rangle$, $|U_B\rangle$, 
$|U_C\rangle$, and $|L\rangle$, it is helpful to use the states
\begin{eqnarray}
\label{mp1}
          &&|1\rangle = 
          3^{-1/2}
          \left(
          |U_A\rangle + |U_B\rangle + |U_C\rangle \right) ,
\\
\label{mp2}
          &&|2\rangle = 
          6^{-1/2}
          \left[
          2|U_A\rangle - \left(|U_B\rangle + |U_C\rangle\right) \right] ,
\\
\label{mp3}
          &&|3\rangle = 
          2^{-1/2}
          \left(
          |U_B\rangle - |U_C\rangle \right) ,
\end{eqnarray}
and $|L\rangle$ as basis states. Note that the states $|1\rangle$
and $|2\rangle$ belong to the $W$ class of the tripartite
entangled states \cite{Dur00}.
If one assumes that the three atoms are identical and equivalently
positioned near the microsphere so that 
\begin{eqnarray}
\label{mp4a}
          &&K_{II}(t-t')=K(t-t'),\quad I=A,B,C ,
\\\label{mp4}
          &&K_{AB}(t-t')=K_{BC}(t-t')=K_{CA}(t-t') ,
\end{eqnarray}
the integro-differential equations for the amplitudes of the states
$|i\rangle|\{0\}\rangle$, \mbox{$i$ $\!=$ $\!1,2,3$} 
decouple [cf. equation~(\ref{e8.5})]:
\begin{eqnarray}
\label{mp5}
&&\dot{C}_1 = \int_0^t \!{\rm d}t'\,
[K(t-t')+2K_{AB}(t-t')] C_1(t') + f_1(t),
\\[.5ex]
\label{mp6}
&&\dot{C}_{2(3)} = \int_0^t {\rm d}t'\, K_-(t-t') C_{2(3)}(t')
+ f_{2(3)}(t),
\end{eqnarray}
where
\begin{eqnarray}
\label{mp8}
       &&f_1(t) =
       3^{-1/2}
       \left[ f_A(t)+f_B(t)+f_C(t)\right]\! ,
\\
\label{mp9}
       &&f_2(t) =
       6^{-1/2}
       \left\{ 2f_A(t) - 
       \left[f_B(t)+f_C(t)\right] \right\} \!,
\\
\label{mp10}
       &&f_3(t) =
       2^{-1/2}
       \left[ f_B(t) - f_C(t)\right] \!.
\end{eqnarray}

In the weak-coupling regime, on applying the Markov approximation,
equations~(\ref{mp5}) and (\ref{mp6}) can easily be solved to obtain
\begin{eqnarray}
\label{mp11}
       &&C_i(t)=\rme^{(-\Gamma_i/2+\rmi\delta_i)t} C_i(0), \quad i=1,2,3,
\\
\label{mp12}
       &&\Gamma_1=\Gamma+2\Gamma_{AB},
       \ \delta_1= \delta+2\delta_{AB} ,
\\
\label{mp13}
       &&\Gamma_2=\Gamma_3=\Gamma_ -,
       \ \delta_2= \delta_3=\delta _-.
\end{eqnarray}
Here we have assumed that the excitation initially resides in the
atomic subsystem. Suppose that atom $A$ is excited at
\mbox{$t$ $\!=$ $\!0$}, then
\mbox{$C_1(0)$ $\!=$ $\!1/\sqrt{3}$}, \mbox{$C_2(0)$ $\!=$ $\!\sqrt{2/3}$},
and \mbox{$C_3(0)$ $\!=$ $\!0$}, and it follows
from equation~(\ref{mp11}) that \mbox{$C_3(t)$ $\!=$ $\!0$}.
If one can set up the system in 
such a way that \mbox{$\Gamma_{AB}$ $\!\sim$ $\!-\Gamma/2$}, then the 
state $|2\rangle$ decays fast, leaving the atomic subsystem in
a mixed entangled state with the weight of  
$|1\rangle\langle 1|$ being less than $1/3$. Alternatively, if 
\mbox{$\Gamma_{AB}$ $\!\sim$ $\!\Gamma$}, then 
the state $|1\rangle$ decays fast,
and the atomic subsystem is prepared in a mixed entangled state 
with the weight of $|2\rangle\langle 2|$ being less than $2/3$. 

If the strong-coupling regime is realized for the state $|1\rangle$, and
if the excitation is pumped into the system through the medium-assisted 
electromagnetic field in such a  way that 
\begin{equation}
\label{mp14}
      f_A(t)=f_B(t)=f_C(t), 
\end{equation}
then \mbox{$f_2(t)$ $\!=$ $\!f_3(t)=0$} [see equations~(\ref{mp9})
and (\ref{mp10})], and
an entangled state of the form ({\ref{e8.16a}), 
with $|1\rangle$ and $C_1(t)$ replacing $|\pm\rangle$
and $C_\pm(t)$, will be generated. 
The condition (\ref{mp14}) can be fulfilled by, e.g., coupling the field first 
to an excited atom D placed at equidistance from the atoms $A$, $B$, and 
$C$. 
In the same way, more than three atoms can be entangled with each other.


\section{Violation of Bell's inequality}
\label{subsec:Bell_ineq}

It is well known that  entangled pairs of spatially separated
particles contradict classical local realism -- the idea that
the properties of one particle cannot be affected instantaneously
by a measurement made upon the other one that is sufficiently
far away. This has been quantified in the form of Bell's inequalities,
which must be obeyed by any local realistic theory, but can be 
violated by quantum mechanics. Whereas violation of Bell's
inequalities always indicates entanglement, the opposite
conclusion is wrong; entanglement must not necessarily
lead to violation of Bell's inequalities. It is therefore
interesting to know the conditions under which violation
of a Bell inequality could principally be observed.
For spin system, the Bell inequality \cite{Bell65}
\begin{equation}
\label{e8.23}
       B_{\rm S}=\big|E(\theta_1,\theta_2)-E(\theta_1,\theta'_2)
       +E(\theta'_1,\theta_2)+E(\theta'_1,\theta'_2)\big|
           \le 2
\end{equation}
is commonly considered, where
\begin{eqnarray}
\label{e8.24}
&\displaystyle
       E(\theta_1,\theta_2) = 
       \bigl\langle \hat{\sigma}_A^{\theta_1} 
       \hat{\sigma}_B^{\theta_2} \bigr\rangle,      
\\[.5ex]
&\displaystyle
\label{e8.25}
       \hat{\sigma}_A^{\theta} = 
       \cos\theta \,\hat{\sigma}_A^x
       + \sin\theta \,\hat{\sigma}_A^y .
\end{eqnarray}
When the state with both atoms simultaneously excited 
is not populated, as 
it is the case for a state of the type (\ref{e8.16a}), it is
not difficult to prove that
\begin{equation}
\label{e8.25a}
        E(\theta_1,\theta_2) = E(\theta_1-\theta_2,0).
\end{equation}
Let us choose
\begin{equation}
\label{e8.25b}
    \theta = \theta_1 - \theta_2 =
    \theta_2 - \theta'_1 = \theta'_1 - \theta'_2\,.
\end{equation}
The inequality (\ref{e8.23}) then simplifies to
\begin{equation}
\label{e8.26}
       B_{\rm S}=|3E(\theta,0) - E(3\theta,0)|
           \le 2 .
\end{equation}

\begin{figure}[tb]
\hspace*{2.5cm}
\psfig{figure=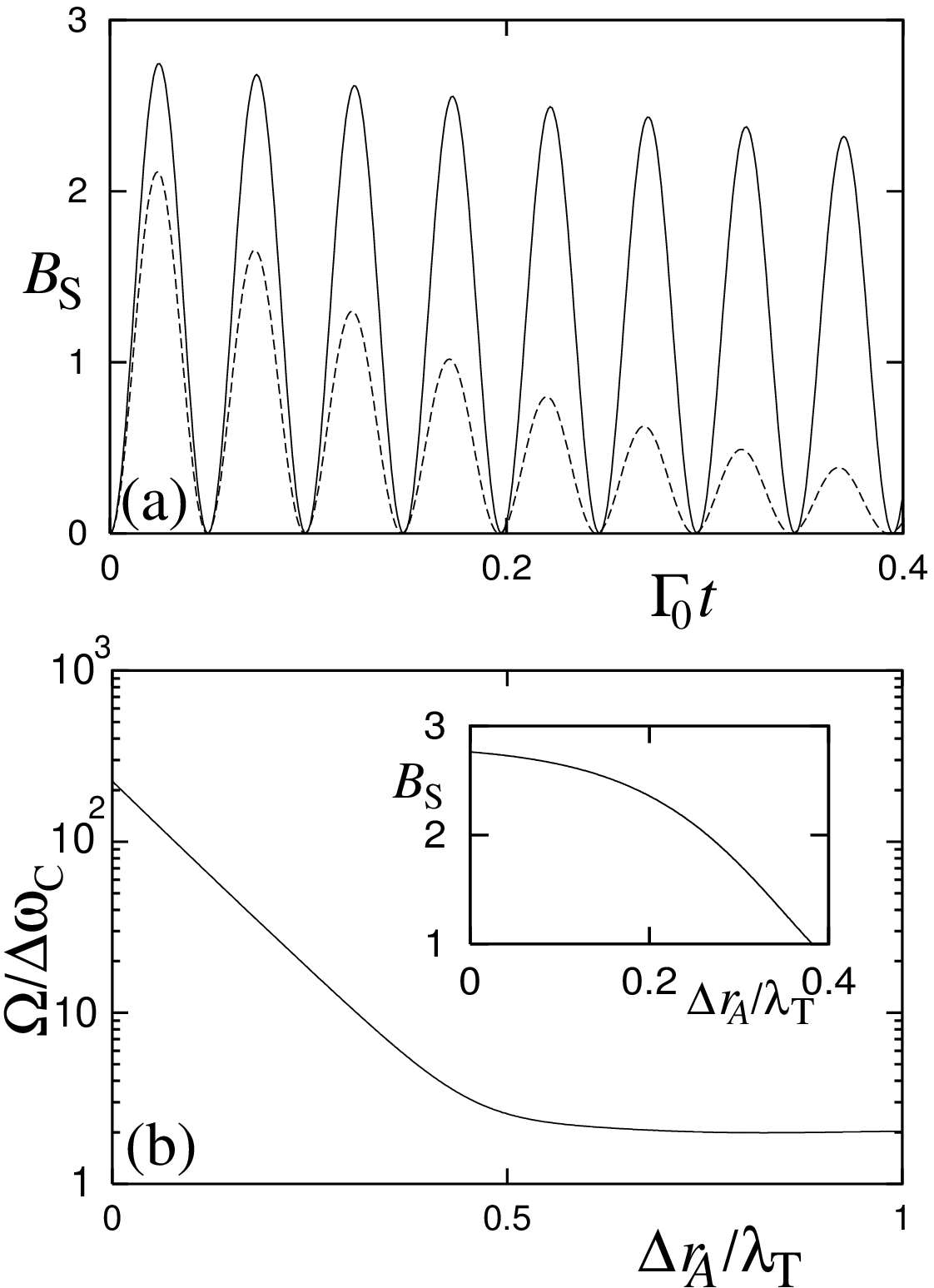,width=.83\textwidth} 
\caption{
The dependence on time of $B_{\rm S}$ is shown for
two atoms at (with respect to a microsphere) diametrically
opposite positions, radially oriented transition dipole
moments, and a single-resonance Drude--Lorentz-type
dielectric
[\mbox{$R$ $\!=$ $\!10\,\lambda_{\rm T}$};
\mbox{$\omega_{\rm P}$ $\!=$ $\!0.5\,\omega_{\rm T}$};
\mbox{$\Delta r_B$ $\!=$ $\!\Delta r_A$ $\!=$ $\!0.02\,\lambda_{\rm T}$};
\mbox{$\omega_A$ $\!=$ $\!1.0501\,\omega_{\rm T}$};
\mbox{$\Gamma_0$ $\!=$ $\!10^{-6}\,\omega_{\rm T}$}; 
$\Omega_D$ $\!=$ $\!\Omega$;
\mbox{$\gamma/\omega_{\rm T}$ $\!=$ $\!10^{-6}$} (solid line),
$10^{-5}$ (dashed line)].
(b) $\Omega/\Delta\omega_{\rm C}$ versus $\Delta r_A$
for \mbox{$\gamma/\omega_{\rm T}$ $\!=$ $\!10^{-6}$}
(\mbox{$\Delta r_A$ $\!\ge$ $\!10^{-3}\,\lambda_{\rm T}$}).
The inset shows the 
variation of the first maximum value of $B_{\rm S}$ in (a). 
}
\label{fig:Bell_ineq}
\end{figure}
An entangled state of the type (\ref{e8.16a}) can 
only give rise to a violation of the Bell's inequality if
\mbox{$|C_\pm(t)|^2$ $\!\ge$ $\!2^{-1/2}$ $\!\simeq$ $\!0.71$}
\cite{Beige00}, which cannot be achieved in the
weak-coupling regime, equation~(\ref{e8.16d}). However, it can be
achieved in the strong-coupling regime [equation~(\ref{e8.21})], where
\begin{eqnarray}
\label{e8.27}
       E(\theta,0) = \pm\cos\theta\,|C_\pm(t)|^2
       = \pm\cos\theta\,
        \rme^{-\Delta\omega_{\rm C}(t+\pi/\Omega_D)}
          \sin^2\left({\Omega t/ \sqrt{2}}\right).
\end{eqnarray}
Substitution of this expression into equation~(\ref{e8.26}) yields, 
on choosing \mbox{$\theta$ $\!=$ $\!\pi/4$}, 
\begin{equation}
\label{e8.28}
       B_{\rm S} = 2\sqrt{2}\,\rme^{-\Delta\omega_{\rm C}(t+\pi/\Omega_D)}
       \sin^2\left({\Omega t/\sqrt{2}}\right),
\end{equation}
which clearly shows that $B_{\rm S}$ $\!>$ $\!2$ becomes possible
as long as \mbox{$\Delta\omega_{\rm C}(t$ $\!+$ $\!\pi/\Omega_D)$
$\!\ll$ $\!1$}.
Examples of the temporal evolution of $B_{\rm S}$ are shown in 
figure~\ref{fig:Bell_ineq}(a).
In figure~\ref{fig:Bell_ineq}(b) the dependence of 
the ratio $\Omega/\Delta\omega_{\rm C}$
on the distance of the atoms from the sphere is plotted. The
strong-coupling regime can be observed for
distances for which \mbox{$\Omega/\Delta\omega_{\rm C}$ $\!\gg$ $\!1$}
is valid. The inset reveals that the maximum value of $B_{\rm S}$ 
decreases with increasing atom-surface distance
and reduces below the threshold value of $2$ still 
in the strong-coupling regime.

It is well known that Bell's inequality tests may
suffer from low detection efficiencies
or from distances between the two entangled parts which are
smaller than the speed of light times 
the measurement time, thus allowing for the two particles
to be connected by a signal during the measurement.
In both cases, from the set of measured data it may
be difficult to decide whether a Bell inequality is really
violated or not. Following \cite{Beige00}, the first loophole
can be closed for massive particle entanglement considered here.
So, the correlation function $E(\theta,0)$ can be determined
experimentally by first applying laser pulses with appropriately 
chosen phases (single qubit rotations) on each of 
the two atoms and then measuring their state populations,
which can be performed with extremely high efficiency.
The second loophole, which typically occurs in Bell's inequality
tests for massive particle entanglement, has not fully
been closed so far \cite{Rowe01,Beige00}.   
The scheme proposed here for particle entanglement can be
realized using, e.g., atomic beams passing nearby a microsphere
or atoms (including quantum dots or other artifical atoms)
at fixed positions. In particular, atomic-beam experiments
might offer a possibility to close the light-cone loophole.


\section{Summary and Conclusions}

The spontaneous emission and the mutual dipole-dipole coupling
of (two-level) atoms can drastically change due to the presence
of macroscopic bodies. The effect can be used to entangle the
atoms with each other. Apart from the shape of the bodies,
their dispersive and absorptive properties are crucial. 
All these aspects can be
taken into account in a consistent way by quantization of the
phenomenological Maxwell field by means of a source-quantity
representation of the
field in terms of the (classical) Green tensor and appropriately
chosen fundamental variables that describe collective excitations
of the field and the matter including the reservoir.
Basically, all that is needed is knowledge about the spatially varying,
complex permittivity of the equipment. As functions of frequency,
the real and imaginary parts of the permittivity must satisfy
the Kramers-Kronig relations, which establish the fundamental
relation between dispersion and absorption.  

The case of two atoms located near a (dispersive and absorptive)
microsphere and single-quantum excitation has been considered
in some detail. It has been shown that
in the weak coupling regime, where the Markov approximation
applies, there is a time window for entangling the atoms.
Entanglement 
up to $0.35\,$ebits can be achieved. The effect is
somewhat unexpected, because it is commonly believed that only strong
atom-cavity coupling can lead to interatomic entanglement.
As shown, in the strong-coupling regime the created
entanglement can be indeed much higher. However, perfect
entanglement (in the sense of a pure Bell state) cannot be achieved
in practice even in the strong-coupling regime.
It is worth noting that Bell's inequality can only be violated
in the strong-coupling regime.

Needless to say that the formalism also applies to the study
of the influence of other types of microcavities on the resonant
atom-light interaction. Throughout the paper 
we have assumed that the mutual distance between the atom
is large enough to disregard the interatomic 
Coulomb interaction. For sufficiently small distances
this approximation fails. Moreover, to include in the theory
the direct short-distance interaction between the atoms,    
the rotating-wave approximation may also fail.
Both problems deserve further investigation.

\ack

We would like to thank C Raabe for some numerical data. This work was
supported by the Deutsche Forschungsgemeinschaft.

}

\end{document}